\documentclass[prl,twocolumn,groupedaddress,showpacs]{revtex4}
\usepackage{graphicx}
\begin{document}
\title{Combustion Models in Finance}

\author{C. Tannous and A. Fessant}
\email{tannous@univ-brest.fr}
\affiliation{Laboratoire de Magnétisme de Bretagne, UMR-CNRS 6135,\\
Université de Bretagne Occidentale, BP: 809 Brest CEDEX, 29285 FRANCE
}
\date{\today}
\begin{abstract}
Combustion reaction kinetics models are used for
the description of a special class of bursty Financial Time Series. 
The small number of parameters they depend upon enable financial analysts to predict the time as well as the magnitude of the jump of the value of the portfolio. Several Financial Time Series are analysed within this framework and applications are given.
\end{abstract}
 
\maketitle
\section{Introduction}

Stock exchange behavior is traditionally analysed with Statistical
tools (Descriptive statistics, Time-series...)  and more recently 
with models derived from High-Energy physics or Statistical Physics.
The importance of financial stakes suffices to justify interest
in these new methods.\\ 

Moreover, any possible analogy that might be drawn with Cooperative
physical phenomena involving a large number of degrees of freedom 
is favourably welcome in the burgeoning field of Financial 
Physics.  The unambiguous identification of the precursory patterns or
aftershock signatures of the market are some of the very important
questions to deal with. It is noted, however, that the US market is   
the favourite candidate analysed in detail so far in the literature 
and a deeper study of the European market scene is lacking, in particular,
the  analysis of individual stock behavior.\\

In some cases, the stock trend is easy to guess, however the largest 
value its rate might attain is very difficult to predict, because of 
the interplay of many parameters drawn from political or economical
interests.\\
   
Obviously, the estimation of this figure is absolutely essential  
for investors and any prediction tool in this field provides a leading
edge to business people who are constantly tapping the market searching
for opportunities of growth and profit. \\ 

We treat this problem by exploiting an analogy with combustion models  
occurring in Physics and Engineering. Surprisingly, the application of 
combustion  theory to stocks picked from various economical and 
manufacturing sectors leads to very realistic estimations close to
current trading values.\\

In order to build a framework for our work, we describe a special class 
of Financial Time Series (FTS) that display bursty behaviour at some 
instant of time with a large jump with respect to a prior stable level of 
activity lasting for a comparatively long time before the burst.\\

The interest in this behaviour for Market Analysts and Financial Companies is to be
able to predict both the time at which the value jumps as well as the amplitude of
the jump in order to assess the magnitude of
benefit or loss in spite of a stable level of activity for some time. The
ability to detect such behaviour provides a new speculation tool for
gauging the potential of some companies and possibly predict the maximum amplitude
of their growth in the near or far future.\\

Models borrowed from Combustion theory display spectacular behaviour
of that sort with a very low pre-ignition state for a long time
followed by a surprising explosion. The resulting behaviour looks
dramatically like the FTS we are interested in.
These Deterministic models are based on the assumption that some chemical
concentration behaves as a disturbance of the combustion kinetics
and is responsible for the sudden explosion with a long somehow
latent pre-ignition state.\\

This modeling can cope with upward or downward bursts with the
proviso of a prior stable activity for a while before the burst.
For instance, our approach cannot cope with situations like 
the recent crisis of the NASDAQ ~\cite{johan}. Market shares and startup
companies  belonging to the "New Economy" belong to a different
class of FTS and their analysis is suggested by Time series analysis
along the lines of the work of Johansen and Sornette ~\cite{johan}
by Stochastic Ito Calculus ~\cite {gardiner} or with Statistical/Field Theory
techniques. Nevertheless, our Deterministic approach embodies patterns of
behaviuor that are akin to what is observed in Cooperative phenomena based
modelling.\\ 

This paper is organised according to the following: In the next section, we describe 
a set of Combustion models that display bursty behaviour with a previously
long pre-ignition state. We extend these models to downward explosive models
and discuss the features of these models. In section 3, we discuss the 
optimisation procedure and the objective function that will allow us to 
predict the burst time and magnitude of the explosion of the FTS. Section
4 reports on the application of these models to actual financial series and
section 5 contains a discussion of the results with our conclusion.
   
\section{Combustion models}
We consider a simple combustion model described by the concentration of a
chemical $y(t)$ that obeys the non-linear evolution equation:
\begin{equation}
\frac{dy}{dt}={y^2}(1-y)
\end{equation}

The initial concentration $y(0)=\epsilon$.\\

This differential equation can be analytically integrated as:

\begin{equation}
ln(\frac{y}{1-y})-\frac{1}{y}=t+C
\end{equation}

where $C=ln(\frac{\epsilon}{1-\epsilon})-\frac{1}{\epsilon}$ is a constant
defined from the initial condition.\\

The time $t^*$ at which the value jumps is defined as the time the curvature
of $y(t)$  changes sign, i.e:

\begin{equation}
t^*=ln(2)-3/2-ln(\frac{\epsilon}{1-\epsilon})+\frac{1}{\epsilon}
\end{equation}

This shows the jump time is on the order of $\frac{1}{\epsilon}$ whereas the width
of the transition region around the jump is defined by:

\begin{equation}
\Delta=ln(\frac{y_2(1-y_1)}{y_1(1-y_2)}) +(\frac{1}{y_1}-\frac{1}{y_2})
\end{equation}

with $y_1$ and $y_2$ are the values defining the transition region around 
$t^*$. We relate these coordinates to $\epsilon$ through: $y_1=\phi_1 \epsilon$ and 
$y_2= 1- \phi_2 \epsilon$  with the constraint: $\phi_1+\phi_2 \leq \frac{1}{\epsilon}$

These expressions show that for an initial value $\epsilon$ and in the 
simple symmetric case $\phi_1=\phi_2=\phi$, the width of the transition
region is given by:

\begin{equation}
\Delta=2 ln(\frac{1-\phi\epsilon}{\phi\epsilon}) +\frac{1}{\phi\epsilon}-\frac{1}{1-\phi\epsilon}
\end{equation}
  
\begin{figure}[htbp]
\begin{center}
\scalebox{0.55}{\includegraphics*{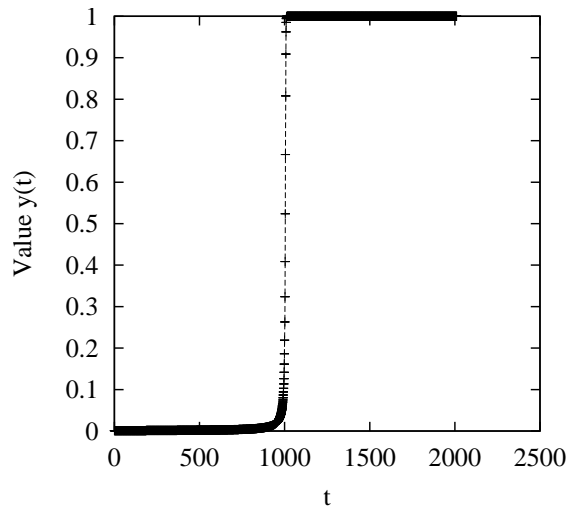}}
\end{center}
\caption{Up burst time series with $\epsilon=10^{-3}$ and $\phi_1=\phi_2=250$. The transition region is bounded by the values [1002,1007]. The unit of time is 1.}
\label{fig1}
\end{figure}

This model is dubbed a Singular perturbation problem ~\cite{omalley} since
decreasing $\epsilon$ induces a divergence of both $t^*$ and $\Delta$.
An example of the behaviour of $y(t)$ is depicted in fig. ~\ref{fig1}
where $\epsilon=10^{-3}$.

The down burst model is described by the differential equation:

\begin{equation}
\frac{dy}{dt}=-y{(1-y)}^2
\end{equation}

with the initial condition $y(0)=1-\epsilon$.

\begin{figure}[htbp]
\begin{center}
\scalebox{0.55}{\includegraphics*{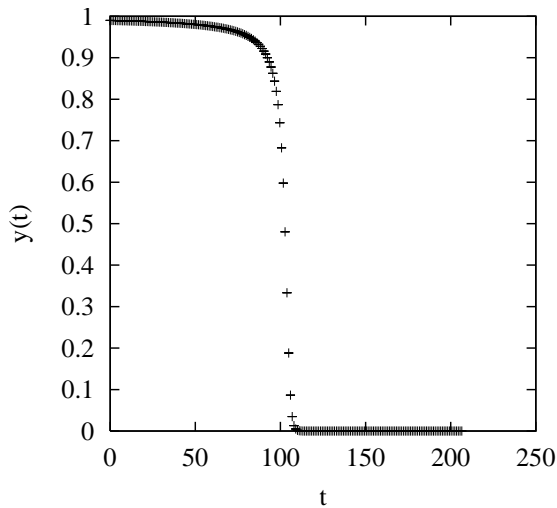}}
\end{center}
\caption{Down burst time series with $\epsilon=10^{-2}$ and $\phi_1=\phi_2=40$. 
The downward transition region is bounded by the values [101.71,103.79] and the unit of time is 1.}
\label{fig2}
\end{figure}

This differential equation can be analytically integrated with the result:
\begin{equation}
ln(\frac{1-y}{y})-\frac{1}{1-y}=t+C
\label{eq1}
\end{equation}

where the constant $C$, the burst time $t^*$ and the transition width are exactly
the same as the previous case if we redefine the values of $y_1$ and $y_2$ in a 
symmetric fashion as: $y_1=1 -\phi \epsilon$ and $y_2= \phi \epsilon$.
The obtained time series for $\epsilon=10^{-2}$ is depicted in fig. ~\ref{fig2}.

\section {Optimisation procedure}

The parameter space consists of three variables: $\epsilon$, $\phi$ and
$\delta t$ the adaptive unit of time. The reason for the existence of
the additional parameter $\delta t$ is that we have to find simulaneously
the best transition time and ratio $\frac{\Delta}{t^*}$ for the FTS.

The optimization procedure consists of defining an objective function
and finding its minimum in the four-dimensional 
parameter space ${\epsilon,\delta t,\phi_1,\phi_2}$.
The objective function is based on a least mean squares approximation of
the difference between the FTS and the Combustion model defined as a functional
$F[\epsilon,\delta t,\phi_1,\phi_2]$.\\

The optimization program itself is based on a globally convergent
method for solving non-linear system of equations: the multidimensional
secant method developed by Broyden ~\cite{broyden65}.
It is based on a fast and accurate method for the iterative evaluation
of the Jacobian of the objective function needed during the minimisation 
procedure. It is a Quasi-Newton method that consists of approximating the Jacobian
and updating it with an iterative procedure. It converges superlinearly
to the solution like all secant methods.\\

In order to run the Optimisation, we tried several strategies based on the
following observations:
\begin{enumerate}
\item The differential equation can be integrated by starting for several trial values
of epsilon and the results for $t^*$ and the ratio $\frac{\Delta}{t^*}$ stored
and interpolated in order to speed up the Optimisation procedure.
\item The analytical solution can be used in order to build the objective
function explicitly, however this is valid only in certain cases and with
the inversion of the roles of y and t enforced by the implicit relation between 
them (see e.g. equation ~\ref{eq1}). 
\item The parameter space being four dimensional was reduced in some cases
to two dimensions by fixing the values of the parameters $\phi_1$ and $\phi_2$
on the basis of statistical grounds. 
\end{enumerate}

All the above operations should yield roughly the same values for
the parameters before running the final check in order to test the
accuracy of the fitting to the FTS at hand. Once the fitting is validated
a prediction for the largest magnitude of the plateau value can be made and 
compared whenever possible to the available data.\\ 

\section {Results}

We have studied the PPS evolution of several companies over a period ranging
from one to five years. These were chosen from different industrial and economic
sectors with a variety of total market capitalizations.
During the same period, the Paris CAC 40 index grew regularly and steadily
except for the July-November 1998 period.\\

These trading shares were chosen because of their peculiar behaviour of underperforming
the CAC 40 index over a period of one to five years with a growth rate that is weaker
than other shares pertaining to the same sector. Some of the shares are
Alcatel, GFI, STMicroelectronics belonging to the TMT sector
(Technology, Media and Telecommunications) whereas DMC, Rochette and Suez
belong to traditional economy sectors. In the Suez case we extended the study 
to seven years.\\

For illustration, we review the cases of these companies one by one highlighting
the validity of the associated combustion model while giving
some historical perspective in order to provide a background 
interpretation of the model parameters.\\ 

\begin{figure}[htbp]
\begin{center}
\scalebox{0.55}{\includegraphics*{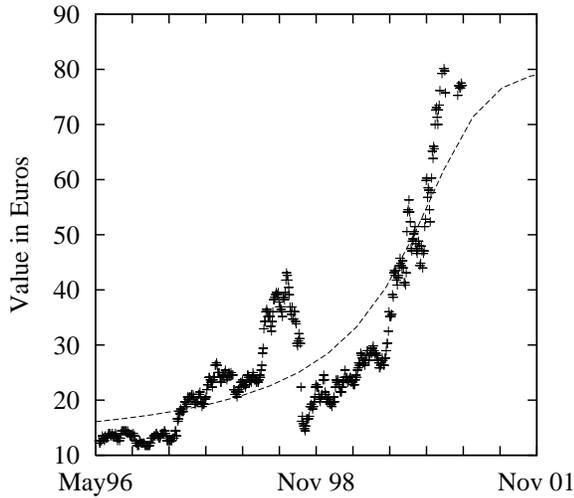}}
\end{center}
\caption{Alcatel time series: The fitting parameters are $\epsilon=0.0796$,
$\delta t=1.4324$, and $\phi_1=\phi_2=2$.}
\label{fig3}
\end{figure}

Starting with Alcatel, its price per share (PPS) started rising, after a stable period of
five years despite a strong dip in November 1997 following the withdrawal
of the American Pension Funds.
The beginning of the growth period spans a period of 2 years with a PPS start
value trading around 20 Euros and a upper value of 90 Euros representing 450 \% growth.
The combustion model predicts an upper value of the PPS trading around 80 Euros which
is an estimation below the actual rate by about 12\% (see fig.~\ref{fig3}).

\begin{figure}[htbp]
\begin{center}
\scalebox{0.55}{\includegraphics*{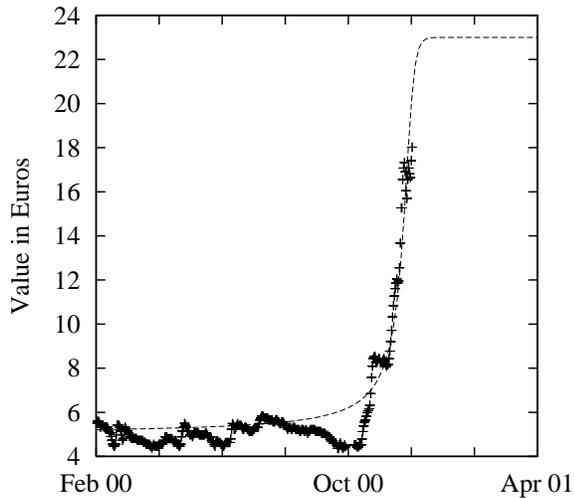}}
\end{center}
\caption{Dolfuss-Mieg time series: The fitting parameters are $\epsilon=0.0118$,
$\delta t=17.922$ and $\phi_1=\phi_2=2$.}
\label{fig4}
\end{figure}

Next is DMC, a textile company that went through difficult times with a serious drop
of its PPS due to a overwhelmingly hostile economic situation. It managed to stabilise its
PPS to trade it at 4 to 5 Euros at the end of 1998.  Ultimately it underwent restructuring and
recentered its activities around Sportswear and Creative Leisure cutting
down on less profitable activities. That action sent a strong signal to stock brokers,
investors and financiers. Its PPS grew 450 \% in less than 2 months to reach a trading value of
about 20 Euros. In this case, the combustion model overestimates the actual
value by 15 \% as displayed in ~\ref{fig4}.

\begin{figure}[htbp]
\begin{center}
\scalebox{0.55}{\includegraphics*{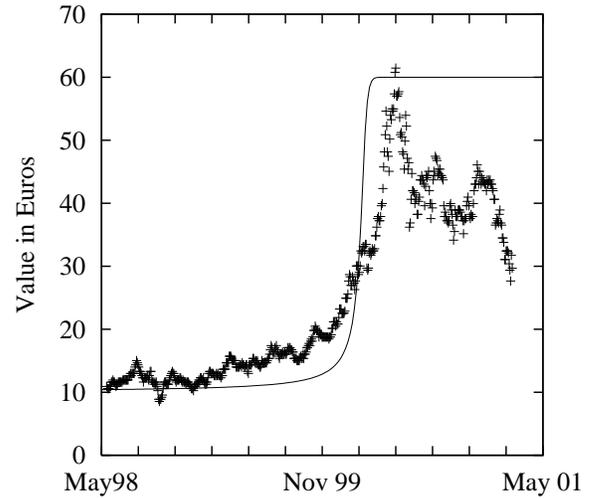}}
\end{center}
\caption{GFI time series: The fitting parameters are $\epsilon=0.0082$,
$\delta t=17.576$ and $\phi_1=\phi_2=2$.}
\label{fig5}
\end{figure}

GFI Informatique is a computer services company belonging to the
TMT sector with obviously a total market capitalisation volume 
much smaller than Alcatel's.
After making its entry on the TMT market in 1998, its PPS started
growing steadily slowly at first then faster around August 1999
due to the interest at that time in TMT companies and excellent prospective
growth potential of GFI. The PPS traded 60 Euros at the beginning of 
the year 2000 reaching a growth rate of 600 \%.
The value predicted by the Combustion model is 60 Euros in perfect 
agreement with the actual value (see fig.~\ref{fig5}).
     
\begin{figure}[htbp]
\begin{center}
\scalebox{0.55}{\includegraphics*{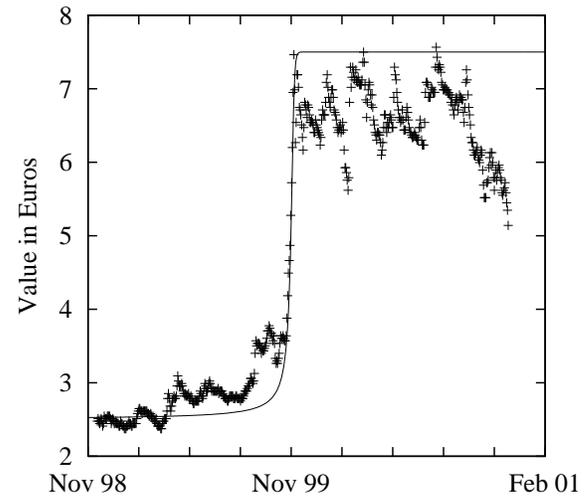}}
\end{center}
\caption{Rochette time series: The fitting parameters are $\epsilon=0.00535$,
$\delta t=47.557$ and $\phi_1=\phi_2=2$.}
\label{fig6}
\end{figure}

Rochette is a traditional Pulp and Paper company with a PPS that suffered
more than 25 \% slip during the general slowdown period of the year 1998
second semester. Its PPS stayed constant for nearly a year trading around
2.5 Euros. At the end of 1999, it grew to 7.7 Euros in good
agreement with the suggested Combustion model (see fig.~\ref{fig6}).

\begin{figure}[htbp]
\begin{center}
\scalebox{0.55}{\includegraphics*{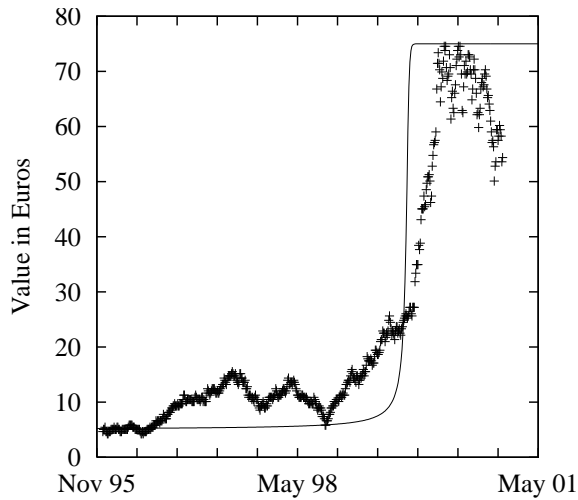}}
\end{center}
\caption{Stmicro time series: The fitting parameters are $\epsilon=0.00317$,
$\delta t=41.268$ and $\phi_1=\phi_2=2$.}
\label{fig7}
\end{figure}

The PPS of STMicroelectronics, a company dealing with the
design and testing of Semiconductor Components had a stable history
until the end of 1998. It exploited fully the development of the High
Technology sector and it rose to 70 Euros and that is equivalent to a 700 \%
progress as predicted by the Combustion model (see fig.~\ref{fig7}).

\begin{figure}[htbp]
\begin{center}
\scalebox{0.55}{\includegraphics*{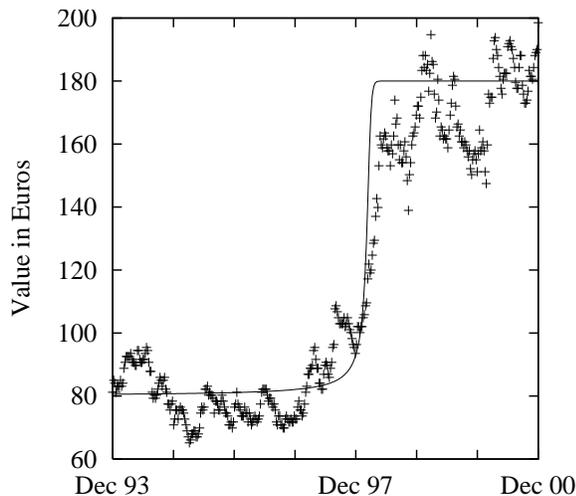}}
\end{center}
\caption{Suez time series: The fitting parameters are $\epsilon=0.00571$,
$\delta t=42.713$ and $\phi_1=\phi_2=2$.}
\label{fig8}
\end{figure}

The final study case is about Suez. Initially, Suez is a company geared toward Energy and Water 
resources distribution; it extended its activity to Telecommunications around 1996
merging with Lyonnaise des Eaux in June 1997. That event triggered a 230 \% growth rate in its
PPS rising from 75 Euros to 175 Euros. Again, this is in with perfect agreement with the 
suggested combustion model (see fig.~\ref{fig8}). The long term stability of the PPS pushed us
in this case to extend the study period to over seven years.

\section{Discussion and Conclusions}

The combustion model is a faithful description of the FTS we considered
and can be used to predict the PPS over some definite period of time ahead.
Is is deterministic and describes features that requires usually the 
mathematics and techniques of Cooperative physical phenomena.
We stress that our study is based on actual PPS time series and not
on synthetic data and that it analyses individual PPS rather than 
composite data that might contain averaging effects.\\

The special class of FTS we consider, suffer some lag spanning periods
that range any time duration from several months to several years with
respect to the progress of other FTS belonging to the same business sector
or to the stock market index.\\

The behaviour we describe seems compatible with the overall picture that
any lag in the progress of the PPS of some company in comparison to other
similar companies makes it attractive to the investors. That interest
even increases as the lag gets more pronounced. In some cases, this induces
a recovery phase in which the PPS readjusts with a rise that is larger 
the more important the respective PPS difference is.\\

The length of the recovery phase depends intimately on the overall 
Economical situation that might interfere with the growth, nevertheless
it is reasonable to expect a readjustment of the price share in such a way
it conforms to the other shares belonging to the same category.\\

The region of validity of the results we obtain spans the stable as well
as the period after the jump of the PPS. We observed that the long term 
evolution of the PPS after the jump is highly variable depending on the
market reaction. In the framework of our model, this means that after the
jump, we are in a stage where the strict conditions for the validity
of the combustion model are no longer obeyed.\\

The simple combustion model is a straightforward translation of the above facts
with the proviso of a sound interpretation of its basic parameters that
ought to be evaluated from the portfolio of the company of interest and
the background Economical context. Identifying the parameters of the combustion
model from Economical data and assessing them is presently work in progress.

\end{document}